\documentclass[fp,twocolumn]{jpsj3}
\UseRawInputEncoding
\usepackage{txfonts}

\title{$S=2$ Quantum Spin Chain with the Biquadratic Exchange Interaction}

\author{T\^oru~Sakai$^{1,2}$\thanks{sakai@spring8.or.jp}, Takaharu~Yamada$^1$%\thanks{jpsj{\_}edit@jps.or.jp}
, Ryosuke~Nakanishi$^1$, Rito~Furuchi$^1$, Hiroki~Nakano$^1$, Hirono~Kaneyasu$^1$, 
Kiyomi~Okamoto$^1$, and Takashi~Tonegawa$^{1,3,4}$ }
\inst{$^1$Graduate School of Science, University of Hyogo, 3-2-1 Kouto, Kamigori, Ako-gun, Hyogo 678-1297, Japan \\
$^2$National Institute for Quantum Science and Technology (QST) SPring-8, Kouto 1-1-1, Sayo, Sayo-gun, Hyogo 679-5148 Japan \\
$^3$Professor Emeritus, Kobe University, Kobe 657-8501, Japan \\
$^4$Department of Physics, Graduate School of Science, Osaka Metropolitan University, Sakai 599-8531, Japan \\
} %\\

\abst{The $S=2$ quantum spin chain with the single-ion anisotropy $D$ and the biquadratic exchange interaction $J_{\rm BQ}$ is investigated 
using the numerical diagonalization of finite-size clusters and the level spectroscopy analysis. 
It is found that the intermediate-$D$ phase corresponding to the symmetry protected topological (SPT) phase 
appears in a wide region of the ground state phase diagram. 
We also obtain the phase diagram at the half of the saturation magnetization
which includes the SPT plateau phase. }

\def\Vec#1{\mbox{\boldmath $#1$}}

\begin{document}
\maketitle

\section{Introduction}

The symmetry protected topological (SPT) phase\cite{gu,pollmann1,pollmann2} has attracted
a great deal of interest in the research field of the 
quantum spin systems and the strongly correlated electron systems.
One of typical examples of the SPT phases is the Haldane phase of the $S=1$ antiferromagnetic chain.\cite{haldane1,haldane2} 
The Haldane phase of the odd-integer $S$ antiferromagnetic chain is the SPT phase, while the even-integer $S$ one is not
(namely, the so-called trivial phase).\cite{pollmann1,pollmann2,capponi,bois,fuji,furuya}
Although the Haldane phase of  the $S=2$ antiferromagnetic chain is not the SPT phase, 
the SPT phase characterized by the string order parameter\cite{dennijs} was predicted\cite{oshikawa}
even in the $S=2$ chain described by
\begin{equation}
   \tilde{\cal H}_0 
   = \sum _{j=1}^L (S_j^x S_{j+1}^x + S_j^y S_{j+1}^y + \lambda S_j^z S_{j+1}^z)
     +D\sum_{j=1}^L (S_j^z)^2,
\label{ham0}
\end{equation}
where $S_j^\mu$ ($\mu = x,y,z$) denotes the $\mu$ component of the $S=2$ spin operator at
the $j$th site,
and $\lambda$ and $D$ are anisotropy parameters.
The numerical diagonalization of finite-size clusters and the level 
spectroscopy analysis\cite{tonegawa, okamoto2011,okamoto2011-2,okamoto2014,okamoto2016} 
applied to this model indicated that 
the SPT phase appeared in a narrow region of the ground state phase diagram on the $\lambda-D$ plane. 
This SPT phase was called the intermediate-$D$ phase.
The earlier density matrix renormalization group (DMRG) calculations\cite{schollwock1, schollwock2, schollwock3} 
applied to the same model could not find the SPT phase, 
probably because the region was too narrow and the energy gap was too small. 
Later the parity DMRG and level spectroscopy analysis successfully indicated the existence of the
SPT phase.\cite{tzeng}
Kj\"all et al.\cite{kjall} also studied the
Hamiltonian (\ref{ham0}) by use of a matrix-product state based DMRG.
They stated that the SPT phase does not exist on the $\lambda-D$ plane and the addition 
of very small positive $D_4 (S_j^z)^4$ brought about the realization of the SPT phase.
This point was discussed in detail by our group.\cite{okamoto2014,okamoto2016}

In any case, the SPT phase, if exists, appears only in such a narrow region of
the ground state phase diagram of the model (\ref{ham0})
that it would be difficult to observe it in any real experiment, 
although the bond-alternation would slightly stabilize the SPT phase.\cite{okamoto2016,ohta}
Several theoretical works suggested that some artificial $S=2$ models  
would exhibit the SPT phase.\cite{okamoto2014,kjall,orus1,orus2,orus3,korepin,mao}

In this paper, we introduce the biquadratic exchange interaction as a more realistic interaction. 
Although the $S=2$ spin chain with the biquadratic interaction had already been 
revealed to exhibit some SPT phases in the previous theoretical works,\cite{orus1,orus2,orus3}
the model also includes the artificial bicubic and biquartic interactions which were fixed to 
special amplitudes suitable for SO(5) symmetry. 
Also it was found by theoretical works\cite{aklt1,aklt2} that the biquadratic interaction enhances the 
Haldane gap of the $S=1$ chain. 
Chiara et al.\cite{chiara} investigated the $S=1$ chain model with the
single-ion anisotropy and the biquadratic exchange interaction to
find the enhancement of the Haldane phase by the latter interaction.
In addition the numerical diagonalization of finite-size clusters and the level spectroscopy analysis 
indicated that this interaction would give rise to the SPT magnetization plateau of the $S=2$ 
antiferromagnetic chain.\cite{sakai}
Thus the biquadratic interaction is expected to stabilize the SPT phase in the 
ground state of the $S=2$ antiferromagnetic chain. 
Using the same analyses as our previous papers,\cite{tonegawa,okamoto2016,sakai}
we investigate the $S=2$ antiferromagnetic 
chain with the biquadratic exchange interaction and the single-ion anisotropy to obtain 
the ground state phase diagram. 
The phase diagram at the half of the saturation magnetization is also presented. 

The remainder of this paper is organized as follows.
We begin with the model description in \S2.
The phase diagrams with the zero magnetization and with the half of the saturation magnetization 
are presented in \S3 and \S4, respectively.
The magnetization curves are given in \S5.
Finally, \S6 is devoted to discussion and \S7 to summary.

\section{Model}

The $S=2$ antiferromagnetic chain with the biquadratic exchange 
interaction and the single-ion anisotropy is investigated. 
The Hamiltonian is given by the following form
\begin{eqnarray}
{\cal H}_0 =\sum _{j=1}^L \Vec{S}_j\cdot \Vec{S}_{j+1} 
+ J_{\rm BQ}\sum_{j=1}^L (\Vec{S}_j\cdot \Vec{S}_{j+1})^2 
+D\sum_{j=1}^L (S_j^z)^2.
\label{ham}
\end{eqnarray}
%The exchange interaction constant is set to be unity as the unit of energy.
For $L$-site systems under the periodic boundary condition ($ \Vec{S}_{L+1}=\Vec{S}_1$), 
the lowest energy of ${\cal H}_0$ in the subspace where 
$\sum _j S_j^z=M$ is denoted as $E(L,M)$. 
The reduced magnetization $m$ is defined by $m=M/M_{\rm s}$, 
where $M_{\rm s}$ denotes the saturation of the magnetization, 
namely $M_{\rm s}=2L$ for the $S=2$ system. 
The Lanczos algorithm is used to calculate $E(L,M)$ up to $L=12$. 
We also calculate the lowest energy under the twisted boundary condition 
($S^{x,y}_{L+1}=-S^{x.y}_1$, $S^z_{L+1}=S^z_1$). 
The eigenstates are distinguished using the space inversion symmetry 
with respect to the twisted bond. 
$E_{\rm TBC+}(L,M)$ denotes the lowest energy with the even parity, 
while $E_{\rm TBC-}(L,M)$ the one with the odd parity.  

\section{Ground State Phase Diagram}

\begin{figure}[h]
   \begin{center}
       \scalebox{0.25}{\includegraphics{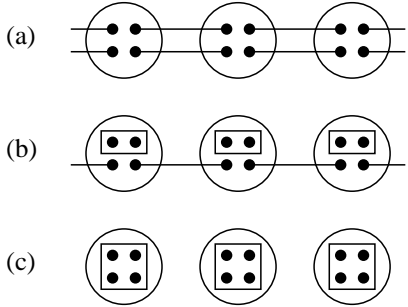}}
   \end{center}
   \caption{VBS pictures for (a) the Haldane state (trivial),  (b) the intermediate-$D$ state (SPT)
   and (c) the large-$D$ state (trivial).
   Big circles denote \hbox{$S=2$} spins and small dots
   \hbox{$S=1/2$} spins.  Solid lines represent valence bonds
   $\bigl($singlet pairs of two \hbox{$S=1/2$} spins,
   $(1/\sqrt{2})(\uparrow\downarrow-\downarrow\uparrow)$$\bigr)$.  Two
   \hbox{$S=1/2$} spins in rectangles are in the
   \hbox{$(S_{\rm tot},S^z_{\rm tot})=(1,0)$} state and similarly four
   \hbox{$S=1/2$} spins in squares are in the
   \hbox{$(S_{\rm tot},S^z_{\rm tot})=(2,0)$} state.}
   \label{fig:vbs-pictures-M=0}
\end{figure}

At first the ground state phase diagram with $m=0$ is investigated. 
According to the previous results,\cite{tonegawa,okamoto2011}
the three gapped phases; (a) the Haldane, (b) the intermediate-$D$, and 
(c) the large-$D$ phases are expected to appear, as well as the 
gapless $XY$ phase in which the spin correlation function 
exhibits the power-law decay. 
We note that the Haldane phase and the large-$D$ phase belong to the same
phase as was shown by our group.\cite{tonegawa,okamoto2011,okamoto2011-2}
The schematic valence bond solid (VBS) pictures of these gapped states are shown in 
%Figs \ref{vbs-pictures-m0} (a), (b) and (c), respectively. 
Figs. \ref{fig:vbs-pictures-M=0} (a), (b) and (c), respectively. 

\begin{figure}
\centerline{\includegraphics[width=0.85\linewidth,angle=0]{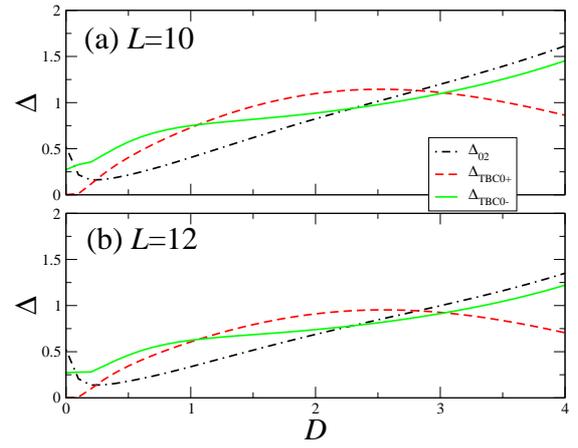}}%
\caption{\label{LSm0} (Color online)
Three gaps, $\Delta_{02}$ (black dashed-dotted line), $\Delta_{\rm TBC+}$ (red broken line)
and $\Delta_{\rm TBC-}$ (green solid line), for $J_{\rm BQ} = 0.12$ 
are plotted versus $D$ for (a) $L$=10 and (b) $L=12$, respectively. 
}
\end{figure}

\begin{figure}
\bigskip
\centerline{\includegraphics[width=0.85\linewidth,angle=0]{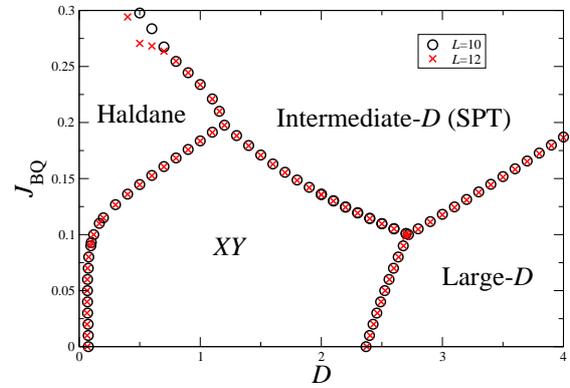}}%
\caption{\label{phasem0} (Color online)
Ground state phase diagram with respect to the single-ion anisotropy $D$ and 
the biquadratic exchange interaction $J_{\rm BQ}$. 
Black circles and red crosses give the phase boundaries estimated 
for $L=$10 and 12, respectively. 
In the large $J_{\rm BQ}$ region ($J_{\rm BQ}\gtrsim 0.25$) 
the system size dependence is too large to determine the phase boundary. 
}
\end{figure}

In order to distinguish these three phases, the level spectroscopy analysis
\cite{kitazawa,nomura-kitazawa} 
is one of the best methods. 
According to this analysis, 
we should compare the following three energy gaps; 
\begin{eqnarray}
\label{delta2}
&&\Delta _{02} =E(L,2)-E(L,0), \\ 
\label{tbc+}
&&\Delta_{\rm TBC0+}=E_{\rm TBC +}(L,0)-E(L,0), \\
\label{tbc-}
&&\Delta_{\rm TBC0-}=E_{\rm TBC -}(L,0)-E(L,0).
\end{eqnarray}
The level spectroscopy method indicates that 
the phase of the ground state can be known from the smallest gap among the above 
three gaps.
Namely, if $\Delta _{02}$ is the smallest,
the ground state is the $XY$ phase.
If $\Delta_{\rm TBC0+}$ is the smallest,
the ground state is the Haldane phase or the large-$D$ phase.
We note that the Haldane phase and the large-$D$ phase
belong to the same phase as already stated.
Finally, if $\Delta_{\rm TBC0-}$ is the smallest,
the ground state is the intermediate-$D$ phase.
The $D$ dependence of these three gaps for $J_{\rm BQ}=0.12$ calculated 
for $L=10$ and 12 is shown in Figs. \ref{LSm0} (a) and (b), respectively. 
$\Delta_{02}$ is the smallest around $D\sim 0$ and $D\sim 4$. 
Thus it is expected that the ground sate is in the Haldane phase around $D\sim 0$, 
while the large-$D$ phase around $D\sim 4$. 
Figure \ref{LSm0} suggests that,
as $D$ increases from 0, 
the quantum phase transition 
from the Haldane phase to the $XY$ phase occurs around $D\sim 2.3$ at first, 
and the one to the large-$D$ phase occurs around $D\sim 3.0$. 
%Each cross point of the gaps indicates the phase boundary. 
These cross points for $L=10$ and 12 are plotted on the plane of $D$ versus $J_{\rm BQ}$ 
in Fig. \ref{phasem0}, as black circles and red crosses, respectively. 
Except for $J_{\rm BQ}\gtrsim 0.25$, the system size dependence of these cross points 
is so small that Fig. \ref{phasem0} is expected to be the ground state phase diagram 
almost in the thermodynamic limit. 
For sufficiently large $J_{\rm BQ}$ the intermediate-$D$ (SPT) phase 
appears in quite wide range of $D$. 
In the large $J_{\rm BQ}$ region ($J_{\rm bQ} \gtrsim 0.25$), 
the system size dependence is too large to estimate the phase boundary, unfortunately. 
Thus it is difficult to conclude whether the Intermediate-$D$ phase can be 
induced only by the biquadratic interaction without $D$ or not.

\section{Magnetization Plateau at the Half of the Saturation Magnetization}

\begin{figure}
\centerline{\includegraphics[width=0.85\linewidth,angle=0]{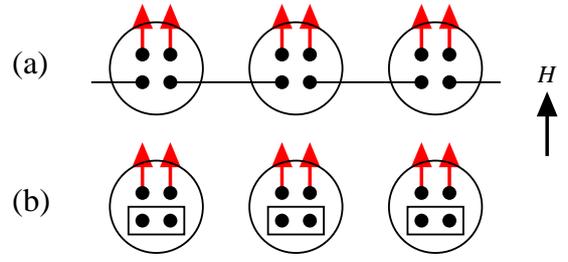}}%
\caption{\label{plateau} (Color online)
Schematic pictures of two mechanisms of the half of the saturation magnetization plateau: 
(a) Haldane plateau and (b) large-$D$ plateau. 
}
\end{figure}

\begin{figure}
\bigskip
\centerline{\includegraphics[width=0.85\linewidth,angle=0]{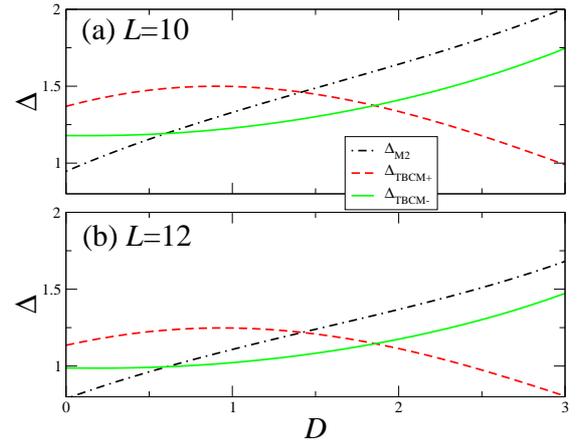}}%
\caption{\label{LSPLA} (Color online)
The $D$ dependence of three gaps, 
$\Delta_{M2}$ (black dashed-dotted line),
$\Delta_{\rm TBCM+}$ (red dashed line) and $\Delta_{\rm TBCM-}$ (green solid line)
 for $J_{\rm BQ} =0.05$ calculated
for (a) $L=$10 and (b) $L=12$. 
}
\end{figure}

\begin{figure}
\bigskip
\centerline{\includegraphics[width=0.85\linewidth,angle=0]{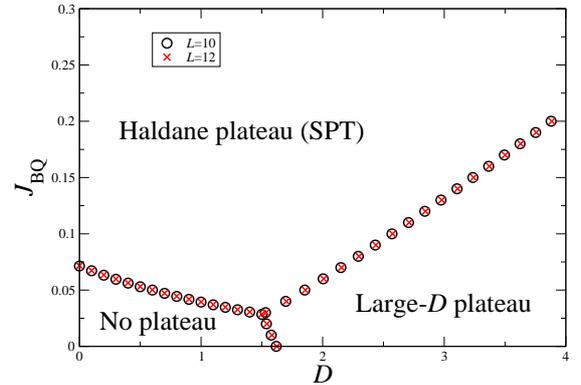}}%
\caption{\label{plateau-phase} (Color online)
Phase diagram at $m=1/2$. Black circles and red crosses are the 
phase boundaries for $L=$10 and 12, respectively, 
determined by the level spectroscopy analysis. 
}
\end{figure}

The SPT phases sometimes appear at the magnetization plateau.\cite{kitazawa-okamoto,sakai,takayoshi}
In the previous work investigating the $S=2$ antiferromagnetic chain 
with the single-ion and the coupling anisotropies,\cite{sakai}
the model was revealed to exhibit the magnetization plateau at the half of the saturation magnetization
based on two different mechanisms, depending on the anisotropies. 
These two mechanisms are the Haldane mechanism and the large-$D$ one, 
which are schematically shown in Figs. \ref{plateau} (a) and (b), 
respectively. 
The Haldane plateau corresponds to the field-induced SPT phase,
while the large-$D$ plateau to the field-induced trivial phase.
This classification is the same as that of the ground state of the $S=1$ chain.
In fact, in Fig.\ref{plateau}, if we remove the small dot spins directing along the magnetic field,
the VBS pictures are nothing but those of the Haldane state and the large-$D$ state of the $S=1$ chain.
The previous work\cite{sakai} indicated that the biquadratic interaction 
stabilizes the Haldane plateau. 
Thus it would be useful to obtain the phase diagram with respect to 
the biquadratic interaction and the single-ion anisotropy. 
For this purpose we investigate the magnetization process of the 
model (\ref{ham}),
by considering the following Hamiltonian
\begin{eqnarray}
\label{magham}
{\cal H}&=&{\cal H}_0+{\cal H}_{\rm Z}, \\
{\cal H}_{\rm Z}&=&-H\sum_jS_j^z,
\label{zeeman}
\end{eqnarray}
where $H$ is the external magnetic field. 
In order to obtain the phase diagram at $m=1/2$, 
the level spectroscopy analysis is also useful. 
According to this method, \cite{kitazawa-okamoto}
we should compare the following three energy gaps; 
\begin{eqnarray}
\label{deltam2}
&&\Delta _{M2} ={E(L,M-2)+E(L,M+2)-2E(L,M) \over 2}, \\
\label{tbcm+}
&&\Delta_{\rm TBCM+}=E_{\rm TBC +}(L,M)-E(L,M), \\
\label{tbcm-}
&&\Delta_{\rm TBCM-}=E_{\rm TBC -}(L,M)-E(L,M),
\end{eqnarray}
where $M$ is fixed to $L$ ($m=1/2$). 
The smallest gap 
among these three gaps determines the phase at $m=1/2$. 
$\Delta_{M2}$, $\Delta_{\rm TBCM+}$ and $\Delta _{\rm TBCM-}$ 
correspond to the gapless (no plateau), the large-$D$-plateau (trivial) and the Haldane-plateau (SPT) phases, 
respectively. 
These three gaps calculated for $L=$10 and 12 for $J_{\rm BQ}=0.05$ 
are plotted versus $D$ in Figs. \ref{LSPLA}(a) and (b), respectively. 
It suggests that as $D$ increases the system at $m=1/2$ changes from the no plateau 
to the Haldane plateau phases, and to the large-$D$ one. 
The phase diagram at $m=1/2$ with respect to $D$ and $J_{\rm BQ}$ is shown 
in Fig. \ref{plateau-phase}, where the level cross points for $L=$10 and 12 
are plotted as black circles and red crosses, respectively. 
The system size dependence of the cross points is so small that 
these points would be the phase boundary almost in the thermodynamic limit. 

\section{Magnetization Curves}

\begin{figure}
\centerline{\includegraphics[width=1.0\linewidth,angle=0]{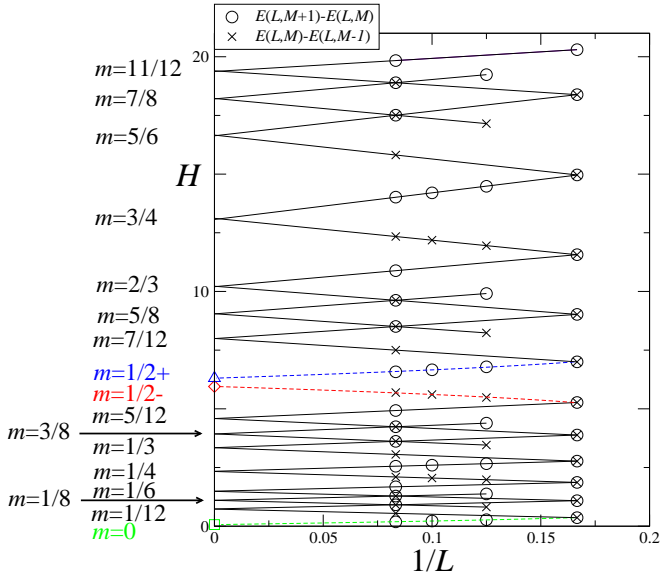}}%
\caption{\label{extramag} (Color online)
$E(L,M+1)-E(L,M)$ and $E(L,M)-E(L,M-1)$ plotted versus $1/L$ with fixed $m$ 
for $J_{\rm BQ}=0.15$ and $D=2.5$. 
Each two quantities seem to coincide with the magnetic field $H$ for $m$ 
in the thermodynamic limit. 
The extrapolated points for $m=0$ (green square), 
$m=1/2-$ (red circle) and 
$m=1/2+$ (blue triangle) correspond to the results of the Shanks transformation.  
Dashed curves are guides for the eye. 
}
\end{figure}

Since the SPT phase and the trivial phase are gapped,
the magnetization gap at $m=0$ and the magnetization plateau at 
$m=1/2$ are expected to appear in the magnetization curve. 
Thus it would be useful to obtain theoretical magnetization curves for 
several typical parameters. 
In order to give the magnetization curve in the thermodynamic limit 
$L \rightarrow \infty$ using the numerical diagonalization results, 
we perform different extrapolation methods in the gapless and gapped cases. 
The magnetic fields $H_-(m)$ and $H_+(m)$ are defined as follows:
\begin{eqnarray}
\label{h1}
E(L,M)-E(L,M-1) \to H_-(m) \quad (L\to \infty), \\
\label{h2}
E(L,M+1)-E(L,M) \to H_+(m) \quad (L\to \infty), 
\end{eqnarray}
where the size $L$ is varied with fixed $m=M/M_{\rm s}$. 
If the system is gapless at $m$, the conformal field theory 
predicts that the size correction is proportional to $1/L$ and 
$H_-(m)$ coincides with $H_+(m)$.\cite{sakai4,sakai5} 
It is justified by Fig. \ref{extramag}, where $E(L,M)-E(L,M-1)$ and 
$E(L,M+1)-E(L,M)$ are plotted versus $1/L$ for
$J_{\rm BQ}=0.15$ and $D=2.5$.
%\lambda =1.0$ and $D=2.0$. 
It suggests that the system is gapless except for $m = 0$ and $1/2$. 
For these magnetizations, we can estimate $H(m)$ in the thermodynamic 
limit, using the following extrapolation form
\begin{eqnarray}
\label{h0}
E(L,M+1)-E(L,M-1) \to H(m) + O(1/L^2).
\end{eqnarray}
On the other hand, if the system has a gap at $m$, namely 
the magnetization plateau is open, $H_-(m)$ does not 
coincides with $H_+(m)$,
and $H_+(m)-H_-(m)$ which corresponds to 
the plateau width. 
In such a case we assume the system is gapped at $m$ and 
use the Shanks transformation\cite{shanks,barber} 
to estimate $H_-(m)$ and $H_+(m)$. 
The Shanks transformation applied to a sequence \{$P_L$\} is 
defined by the form
\begin{eqnarray}
\label{shanks}
P'_{L}={{P_{L-2}P_{L+2}-P_L^2}\over{P_{L-2}+P_{L+2}-2P_L}}.
\end{eqnarray}
As the above level spectroscopy analysis predicts that 
the spin gap at $m=0$ and 
the $m=1/2$ magnetization plateau appear for $J_{\rm BQ}=0.15$ and 
$D$=2.5, we use the method to estimate $H_+(0)$, $H_-(1/2)$ and $H_+(1/2)$.  
For example, the Shanks transformation is applied to the sequence 
$E(L,1/2+1)-E(L,1/2)$ twice to estimate $H_+(1/2)$ (the upper edge of the 1/2 plateau) 
in the thermodynamic limit, as shown in Table \ref{shanks1}. 
Within this analysis 
the best estimation of $H_+(1/2)$ in the thermodynamic limit 
is given by $P''_8$ and the error is determined by the difference 
from $P'_{10}$. 
Thus we conclude $H_+(1/2)=6.3 \pm 0.1$. 

\begin{table}
   \caption{Result of the Shanks transformation applied to the sequence 
   $H_+(1/2)=E(L,1/2+1)-E(L,1/2)]$ twice for $J_{\rm BQ}=0.15$ and $D =2.5$. 
It leads to the conclusion $H_+(1/2)=6.3 \pm 0.1$ in the thermodynamic limit. }
   \bigskip
   \begin{tabular}{|c|c|c|c|}
      \hline
      $L$& $P_L$ & $P_L'$ &$P_L''$ \\ \hline
      ~4~&   7.4209643 & & \\ \hline
      ~6~&   7.0008312  & ~6.5753609~ & \\ \hline
      ~8~&   6.7894387  & 6.4719603 & ~6.3087151~ \\ \hline
     ~10~&   6.6625410 &  6.4086567 & \\ \hline
     ~12~&  6.5779326  && \\ \hline
   \end{tabular}
   \label{shanks1}
\end{table}

\begin{figure}
\centerline{\includegraphics[width=0.85\linewidth,angle=0]{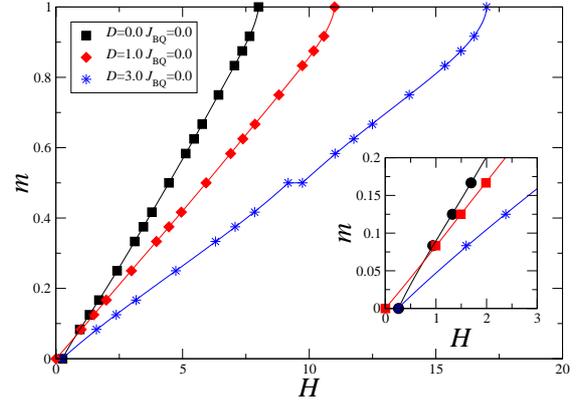}}%
\bigskip
\caption{\label{magbq0} (Color online)
Magnetization curves for $J_{\rm BQ} =0.0$ obtained by the numerical 
diagonalization and the extrapolation methods;  eq.(\ref{h0}) 
for gapless points and the Shanks transformation for plateau points. 
Curves are guides for the eye. 
The Haldane (trivial) gap exists but no plateau for $D=0.0$. 
The magnetization curve for $D=1.0$ has neither the spin gap at $m=0$ nor the plateau at $m=1/2$. 
The large-$D$ (trivial) gap and the large-$D$ (trivial) plateau appear for $D=3.0$. 
Details near the point $(H,m) = (0,0)$ are shown in the inset.}
\end{figure}

\begin{figure}
\vskip1cm
\centerline{\includegraphics[width=0.85\linewidth,angle=0]{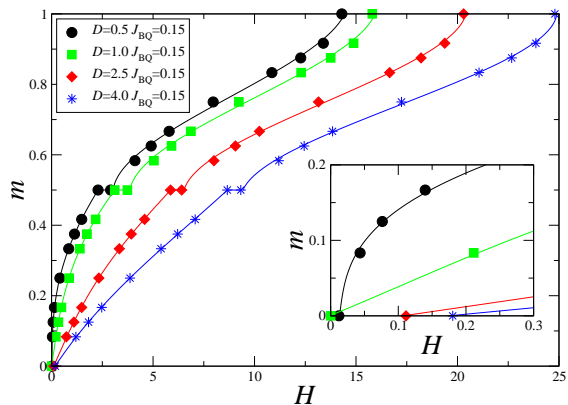}}%
\caption{\label{magbq15} (Color online)
Magnetization curves for $J_{\rm BQ}=0.15$ obtained by the same method 
as Fig. \ref{magbq0}. 
The Haldane (trivial) gap and the Haldane (SPT) plateau appear for $D=0.5$. 
No spin gap but the Haldane (SPT) plateau appear for $D=1.0$. 
The intermediate-$D$ (SPT) gap and the Haldane (SPT) plateau appear for $D=2.5$. 
The large-$D$ (trivial) gap and the large-$D$ (trivial) plateau exist for $D=4.0$. 
Details near the point $(H,m) = (0,0)$ are shown in the inset.}

\end{figure}

Using these methods, the magnetization curves in the thermodynamic limit 
are presented for $J_{\rm BQ} =0.0$ ($D= 0.0$, $1.0$ and $3.0$) in Fig. \ref{magbq0} 
and for $J_{\rm BQ}=0.15$ ($D=0.5$, $1.0$, $2.5$ and $4.0$) in Fig. \ref{magbq15}. 
The intermediate-$D$ phase at $m=0$ and the Haldane plateau at $m=1/2$ 
correspond to the SPT phases. 
Among these magnetization curves both SPT phases appear 
for $J_{\rm BQ}=0.15$ and $D=2.5$ in Fig. \ref{magbq15}.

\section{Discussion}

We have obtained the phase diagrams at $m=0$ (Fig.\ref{phasem0}) and $m=1/2$ (Fig.\ref{plateau-phase}).
The intermediate-$D$ phase in Fig.\ref{phasem0} and the Haldane plateau phase in
Fig.\ref{plateau-phase} are the SPT phases.
Pollmann et al.\cite{pollmann1,pollmann2} showed the existence of a SPT phase
if any one of the following three global symmetries is satisfied: (i) the dihedral
group of ƒÎ rotations about two axes among the $x$-, $y$-, and $z$-axes, (ii) the time reversal
symmetry $\Vec{S}_j \rightarrow -\Vec{S}_j$, and (iii) the space inversion
symmetry with respect to a bond. 
It is easy to see that the Hamiltonian without magnetic field (\ref{ham}) satisfies (ii) and (iii),
whereas the Hamiltonian in magnetic field (\ref{magham}) satisfies only (iii).
Thus, the intermediate-$D$ phase at $m=0$ is protected by (ii) and (iii),
while the Haldane plateau phase at $m=1/2$ by only (iii).

It would be important to discuss the possibility of the experimental 
discovery of the SPT phases for the $S=2$ antiferromagnetic chain. 
As can be seen from the phase diagrams in Figs. \ref{phasem0} and \ref{plateau-phase},
the Haldane plateau (SPT) phase at $m=1/2$ is much wider than the intermediate-$D$ phase at $m=0$.
Thus, as far as this model is concerned,
the experimental finding of the SPT phase is easier at $m=1/2$ than at $m=0$.
A similar situation was also found in our previous work\cite{sakai}.

Furthermore, the intermediate-$D$ phase at $m = 0$ is almost
included in the Haldane plateau one at $m = 1/2$,
except for the narrow region along the boundary contacting the large-$D$ phase.
Then if the ground state is in the intermediate-$D$ phase,
the 1/2 magnetization plateau would appear in the magnetization curve.
The candidate materials of the quasi-one-dimensional $S=2$ antiferromagnet are 
as follows: 
MnCl$_3$(bpy)\cite{hagiwara,shinozaki,fishman}, [Mn(hfac)$_2$]$\cdot$({\it o}-Py-V)\cite{yamaguchi,iwasaki}, 
MnF(salen)\cite{cizmar}, and Pb$_2$Mn(VO$_4$)$_2$(OH).\cite{zhang} 
The 1/2 magnetization plateau has not been discovered for those 
candidate materials, yet. 
Thus the detection of the 1/2 magnetization plateau would be 
a good hint to discover the SPT phase of the $S=2$ antiferromagnetic chain at $m=0$.

\section{Summary}

The $S=2$ antiferromagnetic spin chain with the biquadratic exchange 
interaction and the single-ion anisotropy is investigated using the 
numerical diagonalization of finite-size clusters and the level 
spectroscopy method. 
The ground state phase diagram on the $D-J_{\rm BQ}$ plane 
including the intermediate-$D$ (SPT) phase is obtained. 
The phase diagram at $m=1/2$ including the Haldane (SPT) plateau 
phase is also presented. 
The SPT phases are revealed to appear in quite large regions 
in both phase diagrams. 
In addition the magnetization curves for several typical cases 
are presented.

\section*{Acknowledgment}
%\begin{acknowledgment}
%\acknowledgment

This work was partly supported by JSPS KAKENHI, 
Grant Numbers JP16K05419, JP20K03866, JP16H01080 (J-Physics), 
JP18H04330 (J-Physics) and JP20H05274.
A part of the computations was performed using
facilities of the Supercomputer Center,
Institute for Solid State Physics, University of Tokyo,
and the Computer Room, Yukawa Institute for Theoretical Physics,
Kyoto University.
In this research, we used the computational resources 
of the supercomputer Fugaku provided by the RIKEN 
through the HPCI System Research projects (Project ID: 
hp200173, hp210068, hp210127, hp210201, and hp220043). 
%\end{acknowledgment}

\end{document}